\documentclass[twocolumn,epjc3]{svjour3}          

\usepackage{epsfig,amsmath,amssymb}
\usepackage{xcolor}

\newcommand{\Exp}[1]{\langle#1\rangle}

\RequirePackage[T1]{fontenc}

\smartqed  

\RequirePackage{graphicx}
\RequirePackage{mathptmx}      
\RequirePackage{flushend}
\RequirePackage[numbers,sort&compress]{natbib}
\RequirePackage[colorlinks,citecolor=blue,urlcolor=blue,linkcolor=blue]{hyperref}

\journalname{Eur. Phys. J. C}

\begin{document}

\title{Fisher information and the weak equivalence principle of a quantum particle in a gravitational wave}

\author{James Q. Quach\thanksref{e1,addr1}
}

\thankstext{e1}{e-mail: quach.james@gmail.com}

\institute{Institute for Photonics and Advanced Sensing and School of Chemistry and Physics, The University of Adelaide, South Australia 5005, Australia\label{addr1}
}

\date{Received: date / Accepted: date}

\maketitle

\begin{abstract}
	We show that the weak equivalence principle (WEP) is violated for a quantum particle in a gravitational wave (GW) background, in the sense that extra mass information can be extracted in the presence of the GW. We quantify the degree of violation with the Fisher information of mass. This provides a precise characterisation of WEP violation by quantum systems in a GW, that should be useful in formalising other works that have argued for such violations heuristically.
\end{abstract}

\section{Introduction}
\label{sec:Introduction}

The WEP states that point particles in free-fall will follow trajectories that are independent of their mass. This principle underpins classical gravitational theory. In the context of classical theory the WEP is well defined; in quantum theory however, the WEP is ill-defined. This is because under the Heisenberg's uncertainty principle, point particles and trajectories are ambiguous concepts. The problem is further highlighted when one compares the classical action of a particle with mass $m$ in a gravitational field with the quantum action of a wavefunction $\psi$ of a massive spinless particle. The classical action is
\begin{equation}
S_\text{C}=-mc\int ds~,
\end{equation} 
where $ds^2=g_{\mu\nu}dx^\mu dx^\nu$. As $m$ appears simply as a multiplicative factor, it does not feature in the equations of motion. This is consistent with the WEP. In comparison, the quantum action is
\begin{equation}
S_\text{Q}=-\frac{\hbar^2}{2m}\int\sqrt{g}\Big(g^{\mu\nu}D_\mu\Phi^\dagger D_\nu\Phi+\frac{m^2c^2}{\hbar^2}\Phi^\dagger\Phi\Big)d_4x~,
\end{equation}
where $D_\mu$ is the covariant derivative in curved space-time. In this case $m$ simply is not a multiplicative factor, and features in the Klein-Gordon equation.

In this background, there have been suggestions that the properties of \emph{quantum fluids} (superconductors, superfluids, quantum Hall fluids, Bose-Einstein condensates) may enhance the interaction with GW, leading to superfluids as a medium for gravitational antennae~\cite{papini70,anandan81,chiao82,anandan82,anandan84,peng90,peng91}, superconducting circuits as GW detectors~\cite{anandan85}, transducers~\cite{chiao03, licht04} and mirrors~\cite{minter10, quach15,quach15b}. These idea have not been met without controversy~\cite{kowitt94,harris99,keifer05}. The reason for this is that many of these ideas heuristically apply the notion that quantum particles violate the WEP. This motivates us to provide a more rigorous characterisation of the WEP for quantum particles in GWs.

The WEP's notion that free-falling trajectories should be independent of mass, can be reformulated as the statement that the Fisher information of a free-falling object is invariant with mass~\cite{seveso17}. In this information-theoretic framework, violation of the WEP means that one may extract information about an object's mass in free-fall. This information-theoretic formulation of the WEP has the advantage that it is extendable to quantum objects in an unambiguous manner. Specifically, the Fisher information gives the amount of information that an observable random variable provides about an unknown parameter. In our case, the random variable is the position of the particle $\mathbf{x}$, and the unknown parameter is its mass $m$. For a particle with wave function $\psi(\mathbf{x},t)$, the Fisher information is
\begin{equation}
F_x(m)=\int d\mathbf{x}|\psi(\mathbf{x},t)|^2[\partial_m\log|\psi(\mathbf{x},t)|^2]^2~.
\label{eq:fisher}
\end{equation}

In the absence of gravity, observation of the position of the particle can betray information about its mass. For example, a free Gaussian wave packet spreads with variance $\sigma^2(t) = \sigma^2(0) + \hbar t/2m$; one may extract information about its mass by monitoring its position. Formulation of the WEP in terms of Fisher information states that the presence of a  gravitational field should not produce more information about the mass of a particle, \textit{i.e.} $F_x(m)=F_x(m)^\text{free}$, where $F^\text{free}_x(m)$ is the Fisher information in the absence of a gravitational field.  

In Sec.~\ref{sec:Static gravitational field} we look at quantum particles in static uniform gravitational field. In particular we will investigate the time evolution of Gaussian wavepackets in the Schwarzschild metric, and their related mass Fisher information. We will show how the wavepacket follows the WEP in a static uniform field. In Sec.~\ref{sec:GW} we will derive the time evolution of Gaussian wavepackets in a GW background, and investigate their mass Fisher information. We will show how the wavepacket violates the WEP in the time-dependent uniform field.

\section{Static uniform gravitational field}
\label{sec:Static gravitational field}
We investigate whether extra mass information can be extracted from a Gaussian wavepacket in a Schwarsczhild field. We begin with the Dirac equation in curved space-time, which describes a spin-1/2 particle of rest mass $m$ in a gravitational field,
\begin{equation}
i\hbar\gamma^ae^\mu_a(\partial_\mu-\Gamma_\mu)\psi=mc\psi~.
\label{dirac}
\end{equation}
The spacetime metric $g_{\mu\nu}$ can be related at every point to a tangent Minkowski space $\eta_{ab}$ via tetrads $e_\mu^a$, $g_{\mu\nu}=e_\mu^a e_\nu^b\eta_{ab}$. The tetrads obey the orthogonality conditions $e_\mu^a e_a^\nu=\delta_\mu^\nu,e_\mu^a e_b^\mu=\delta_b^a$. We use the convention that Latin indices represent components in the tetrad frame. The spinorial affine connection $\Gamma_\mu=\frac{i}{4} e_\nu^a (\partial_\mu e^{\nu b}+\Gamma_{\mu \sigma}^\nu e^{\sigma b}) \sigma_{ab}$, where $\Gamma_{\mu\sigma}^\nu$ is the affine connection and $\sigma_{ab}\equiv\frac{i}{2}[\gamma_a,\gamma_b]$ are the generators of the Lorentz group. $\gamma_a$ are gamma matrices defining the Clifford algebra $\{\gamma_a,\gamma_b\}=-2\eta_{ab}$, with spacetime metric signature ($-,+,+,+$). We use the Einstein summation convention where repeated indices ($\mu,\nu,\sigma,a,b=\{0,1,2,3\}$) are summed.

We will consider the Schwarzschild metric in isotopic coordinates ($x^0\equiv ct$),
\begin{equation}
ds^2=V^2(dx^0)^2-W^2(d\mathbf{x}\cdot d\mathbf{x})~,
\end{equation}
where ($r\equiv\sqrt{\mathbf{x}\cdot\mathbf{x}}$)
\begin{gather}
V=\Big(1-\frac{GM}{2c^2r}\Big)\Big(1-\frac{GM}{2c^2r}\Big)^{-1}~,\\
W=\Big(1+\frac{GM}{2c^2r}\Big)^2~.
\end{gather}
Under this metric Eq.~(\ref{dirac}) can be written in the familiar Schr\"{o}dinger picture $i\hbar\partial_t \psi=H\psi$, where ($\boldsymbol{\alpha}\equiv\gamma^0\boldsymbol{\gamma},\beta\equiv\gamma^0,\boldsymbol{p}\equiv-i\hbar\nabla, F\equiv V/W$, and indices $i,..,n=\{1,2,3\}$~\cite{obukhov01}
\begin{equation}
H=\beta mc^2V + \frac{c}{2}[(\boldsymbol{\alpha}\cdot\mathbf{p})F-F(\boldsymbol{\alpha}\cdot\mathbf{p})]~.
\label{eq:H}
\end{equation}

A means by which to write down the non-relativistic limit of the Dirac Hamiltonian with relativistic correction terms is provided by the Foldy-Wouthuysen (FW) transformation~\cite{foldy78}. The FW transformation is a unitary transformation which separates the upper and lower spinor components. In the FW representation, the Hamiltonian and all operators are block-diagonal (diagonal in two spinors). There are two variants of the FW transformation known as the \textit{standard} FW (SFW)~\cite{foldy78} and \textit{exact} FW (EFW)~\cite{eriksen58,nikitin98,obukhov01,jentschura14} transformations. We will use here the EFW transformation, which is efficient and correct at low-orders, and adequate for our purposes. For higher-order corrections one should use the more involved SFW, as the EFW may produce spurious results at higher-orders, in some instances~\cite{jentschura13}.

Central to the EFW transformation is the property that when $H$ anti-commutes with $J\equiv i\gamma^5 \beta$, $\{H,J\}=0$, under the unitary transformation $U=U_2 U_1$, where ($\Lambda\equiv H/\sqrt{H^2}$) 
\begin{equation}
U_1=\frac{1}{\sqrt{2}} (1+J\Lambda),\quad\quad U_2=\frac{1}{\sqrt{2}} (1+\beta J)~,
\end{equation}
the transformed Hamiltonian is even (even terms do not mix the upper and lower spinor components, odd terms do), 
\begin{equation}
\begin{split}
UHU^+=&\frac12\beta(\sqrt{H^2}+\beta\sqrt{H^2}\beta)+\frac12(\sqrt{H^2}-\beta\sqrt{H^2}\beta)J\\
=&\{\sqrt{H^2}\}_\text{even}\beta+\{\sqrt{H^2}\}_\text{odd}J~.
\end{split}
\label{UHU}
\end{equation}
Note that as $\beta$ is an even operator and $J$ is an odd operator, Eq.~(\ref{UHU}) is an even expression which does not mix the positive and negative energy states.

Our Hamiltonian satisfies the EFW anti-commutation property. Using the identity $\alpha^i\alpha^{j}=i\epsilon^{ijk}\sigma_k \textbf{I}_{2} + \delta^{ij}\textbf{I}_{4}$, the perturbative expansion of $\sqrt{H^2}$ yields to first order,
\begin{equation}
\begin{split}
H\approx mc^2V + \frac{1}{4m}(W^{-1}p^2F + Fp^2W^{-1})~.
\end{split}
\label{H_FW}
\end{equation}

Note that $\sqrt{H^2}=\{\sqrt{H^2}\}_\text{even}=H \textbf{I}_{2}$ contains only even terms, and therefore $\{\sqrt{H^2}\}_\text{odd}=0$ in Eq.~(\ref{UHU}).

Taking the weak-limit gravitational field limit so that,
\begin{equation}
V \approx 1-\frac{GM}{c^2r}~,\quad  W \approx 1+\frac{GM}{c^2r}~,
\end{equation}
we get ($\mathbf{g}\equiv-GM\mathbf{r}/r^3$)
\begin{equation}
H = mc^2 + \frac{p^2}{2m} + m\mathbf{g}\cdot\mathbf{x}~.
\label{eq:H_G}
\end{equation}

The Dirac equation in curved spacetime will also give rise to a spin-gravity coupling term ($- \frac{\hbar}{2c}\mathbf{\sigma}\cdot\mathbf{g}$), which we neglected beginning at Eq.~(\ref{H_FW}).  Here we will simply look at the mass Fisher information of a Gaussian wavepacket in a gravitational field to first order. We leave the consideration of higher-order spin-gravity coupling terms in further work. 

The evolution of a quantum particle is governed by the time-evolution operator $U = e^{-iHt}$. Taking the Baker-Campbell-Hausdorff expansion of $U$ to second-order, the time-evolution operator in a Schwarzschild field is ($\hbar=1$)~\cite{seveso17}
\begin{equation}
\begin{split}
U \approx& \exp\Big(\frac{imt^3}{3}\mathbf{g}^2\Big)\exp\Big(\frac{it^3}{6m}\nabla\mathbf{g}\cdot\nabla\nabla-\frac{\mathbf{g}t^2}{2}\cdot\nabla\Big)\\
&\exp\Big(-imt\mathbf{g}\cdot\mathbf{x}\Big)U_\text{free}~,
\end{split}
\label{eq:U_full}
\end{equation}
where $U_\text{free}=\exp(-imc^2t)\exp(-it\Delta/2m)$ is the free time-evolution operator in the absence of any gravitational field. Note that the $\exp(-imc^2t)$ term only acts as a constant phase factor in the non-relativistic limit, and therefore can be ignored.

As our gravitational field is spherically symmetric, we can reduce our problem to one spatial dimension in the radial direction. We consider a Gaussian wave packet, 
\begin{equation}
\psi(\mathbf{x},0) = \Big(\frac{2}{\pi}\Big)^{1/4}e^{-r^2}~,
\label{eq:wavepacket}
\end{equation}
as this is most amenable to comparison with a classical particle. For probe particles travelling over small distances, it is usual to take the terrestrial gravitational field as uniform. In the uniform gravitational case,
\begin{equation}
\begin{split}
U =& \exp\Big(\frac{imt^3}{3}\mathbf{g}^2\Big)\exp\Big(-\frac{\mathbf{g}t^2}{2}\cdot\nabla\Big)\\
&\exp\Big(-imt\mathbf{g}\cdot\mathbf{x}\Big)U_\text{free}~.
\end{split}
\end{equation}

Using the fact that the momentum operator is a translation operator in the conjugate position space, the time evolution of the wave function [$\psi(\mathbf{x},t) = U\psi(\mathbf{x},0)$] is
\begin{equation}
\begin{split}
\psi_\text{g}(\mathbf{x},t) =&\exp\Big(\frac{imt^3}{3}\mathbf{g}^2\Big)\exp\Big(-imt\mathbf{g}\cdot\mathbf{x}\Big)\\
&\psi_\text{free}(\mathbf{x}-\frac{\mathbf{g}t^2}{2},t)~,
\end{split}
\label{eq:psi_uniform}
\end{equation}
where $\psi_\text{free}(\mathbf{x},t)=U_\text{free}\psi(\mathbf{x},0)$ is the free wave function in the absence of a gravitational field. The uniform gravitational field induces a mass-dependent phase factor in Eq.~(\ref{eq:psi_uniform}). This mass-dependent phase factor however, is not present in the probability distribution, $|\psi_g(\mathbf{x},t)|^2=|\psi_\text{free}(\mathbf{x}-\mathbf{g}t^2/2,t)|^2$. Therefore, by a change of variable ($\mathbf{u}=\mathbf{x}-\mathbf{g}t^2/2$), we see that the uniform gravitational field does not produce any extra mass information, \textit{i.e.}
\begin{equation}
\begin{split}
F_x^\text{g}(m)&=\int d\mathbf{u}|\psi_\text{free}(\mathbf{u},t)|^2[\partial_m\log|\psi_\text{free}(\mathbf{u},t)|^2]^2\\
&=F_x^\text{free}(m)~.
\end{split}
\end{equation}

The expected position of the wave packet in a uniform gravitational field is
\begin{equation}
\Exp{\mathbf{x_\text{g}}} = \int_{-\infty}^{\infty}\psi_\text{g}(\mathbf{x},t)^*\mathbf{x}\psi_\text{g}(\mathbf{x},t)d\mathbf{x}= \frac{\mathbf{g}t^2}{2}~.
\end{equation}
This is the geodesic of a freely-falling classical particle with no initial momentum in a uniform gravitational field $\mathbf{g}$. As with the classical case, the expected trajectory of the quantum particle is independent of its mass, in alignment with the WEP, and our finding that there is no more mass Fisher information generated in the presence of a static uniform gravitational field.

\section{Time-dependent uniform gravitational field}
\label{sec:GW}

The metric for a generally polarised linear plane GW is
\begin{equation}
ds^2=-c^2dt^2+dz^2+(1-2v)dx^2+(1+2v)dy^2-2udxdy~,
\label{gw_metric}
\end{equation}
where $u=u(t-z)$ and $v=v(t-z)$ are functions which describe a wave propagating in the $z$-direction. We will consider the case of a circularly polarised GW travelling along the $z$-direction, \textit{i.e.} $v=f=f_0\cos(kz-\omega t)$ and $u=if$. Under this metric, Eq.~(\ref{dirac}) can be written in the familiar Schr\"{o}dinger picture as~\cite{goncalves07,quach15a}
\begin{equation}
H=\beta mc^2+c\alpha^j(\delta_j^i+T_j^i)p_i~,
\label{eq:H}
\end{equation}
with
\begin{equation}
T=
\begin{pmatrix}
v& &-u &0\\
-u& &-v &0\\
0& &0 &0\\
\end{pmatrix}~.
\label{eq:T}
\end{equation}

Applying an EFW (or SFW) transformation, as in the previous static uniform gravitational field case, and ignoring higher-order terms, one arrives at~\cite{quach16}
\begin{equation}
\begin{split}
H_\text{GW}=&\frac{1}{2m}(\delta^{ij}+2T^{ij})p_ip_j + mc^2.
\end{split}
\label{H_GW}
\end{equation}

We would like to know how a Gaussian wave packet behaves in a GW background. We will consider the wave packet located at $z=0$, in one spatial dimension $x$, without loss of generality in our conclusions~\cite{note1},
\begin{equation}
\psi(x,0) = \Big(\frac{2}{\pi}\Big)^{1/4}e^{-(x-x_0)^2}~.
\label{eq:wavepacket2}
\end{equation}

We apply the unitary transformation operator\\ $U=\exp[-i/\hbar\int H_\text{GW}(t)dt]$ to Eq.~(\ref{eq:wavepacket2}) to get the time-evolution of a wave packet in a GW background. Specifically, we Fourier transform the wave packet [Eq.~(\ref{eq:wavepacket2})] into momentum space, $\psi(k,0) = \Big(\frac{1}{2\pi}\Big)^{1/4}e^{-k^2/4}e^{-ikx_0}$~. This allows us to easily write down the time-evolution of the wave pack in momentum space,
\begin{equation}
\begin{split}
\psi_\text{gw}(k,t) &= e^{-i/\hbar\int H_\text{GW}(t)dt} \psi(k,0)\\ 
&=\Big(\frac{1}{2\pi}\Big)^{1/4}e^{-i\hbar k^2[t+f_0\sin(\omega t)/\omega]/2m}e^{-k^2/4}e^{-ikx_0}~.
\end{split}
\label{eq:wavepacket_k}
\end{equation}
where we have used the fact that $H_\text{GW}$ in one dimension is $
H_\text{GW}=\frac{1+f_0\cos\omega t}{2m}p^2$~. We Fourier transform back into position space to arrive at
\begin{equation}
\psi_\text{gw}(x,t)=\Big(\frac{2}{\pi}\Big)^{1/4}\frac{e^{-(x-x_0)^2/b}}{\sqrt{b}}
\label{psi_gw}
\end{equation}
\begin{equation}
b\equiv 1+\frac{2i\hbar}{m}[t+f_0\sin(\omega t)/\omega]~.
\label{eq:b}
\end{equation}

\begin{figure}
	\centering
	\includegraphics[width=\columnwidth]{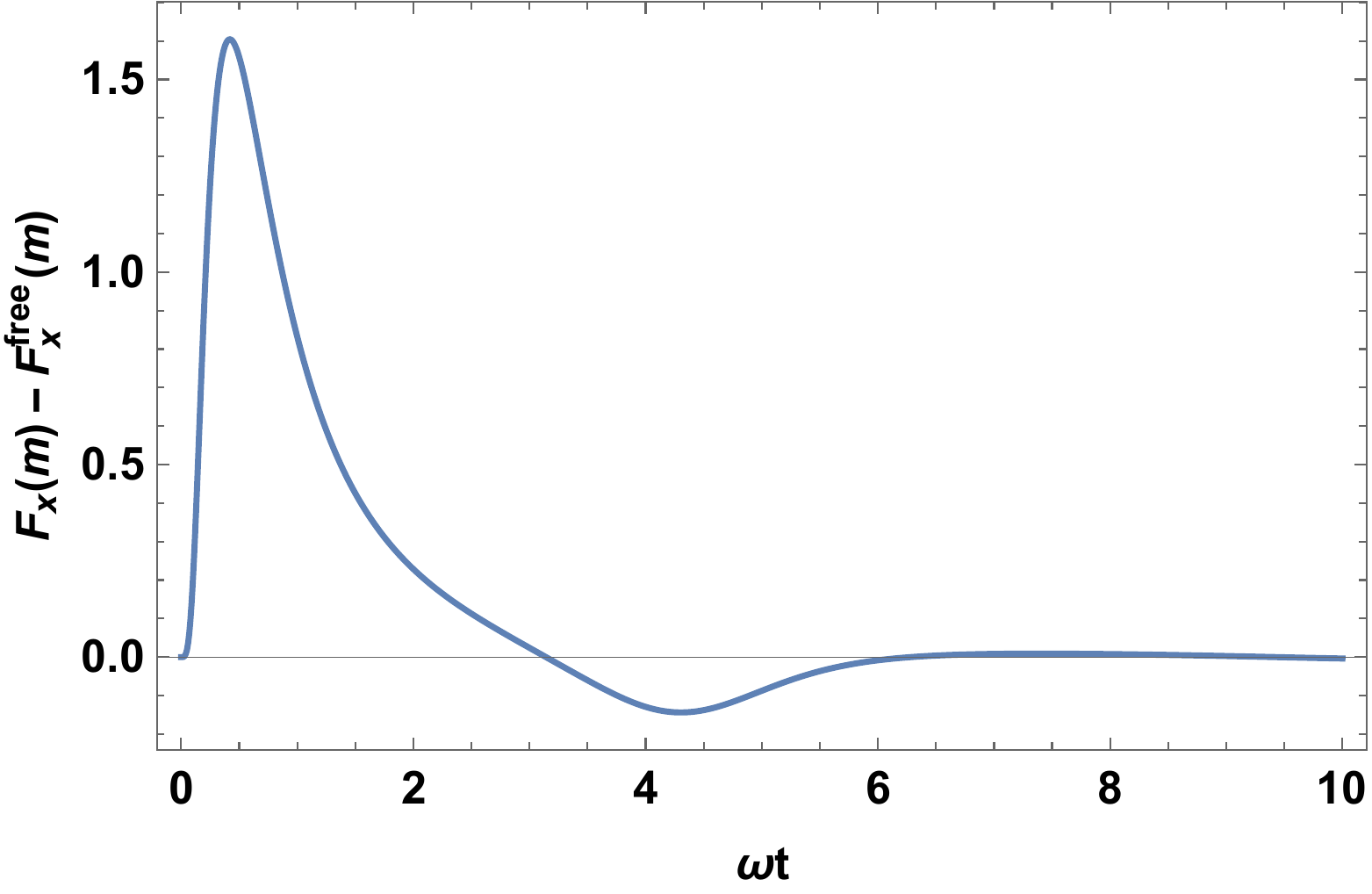}
	\caption{The difference in the mass Fisher information in a GW background from the free case, over time. As the difference can be non-zero, one can extract mass information of the particle from the GW, in violation of the WEP.}
	\label{fig:fisher}
\end{figure}

From Eq.~(\ref{psi_gw}), the expected position of the wave packet in a GW background is
\begin{equation}
\Exp{x_\text{gw}} = x_0~.
\label{eq:wavepacketGW}
\end{equation}
In other words, the particle is expected to remain at rest in the co-ordinate system of Eq.~(\ref{gw_metric}). This actually is not surprising as this is also what happens in the classical case. In the classical case, the presence of a GW is measured in the change of the proper distance between two particles. However, unlike the classical case, the presence of the GW will generate mass information. From Eq.~(\ref{psi_gw}), we numerically calculate the mass Fisher information [Eq.~(\ref{eq:fisher})] of the particle in the GW, and compare it to the free case. To reveal the effects of the GW, $f_0$ and $m$ are set to unity (and $\hbar=1$).  More realistic values are discussed later. Fig.~\ref{fig:fisher} plots the difference in the mass Fisher information in a GW background from the free case, showing that in general it is different from zero. This means that one can extract mass information of the particle from the GW, in violation of the WEP, in stark contrast to the static uniform gravitational field of the previous section, \textit{i.e.}
\begin{equation}
F_x^\text{gw}(m)\neq F_x^\text{free}(m)~.
\end{equation}

\begin{figure}
	\centering
	\includegraphics[width=\columnwidth]{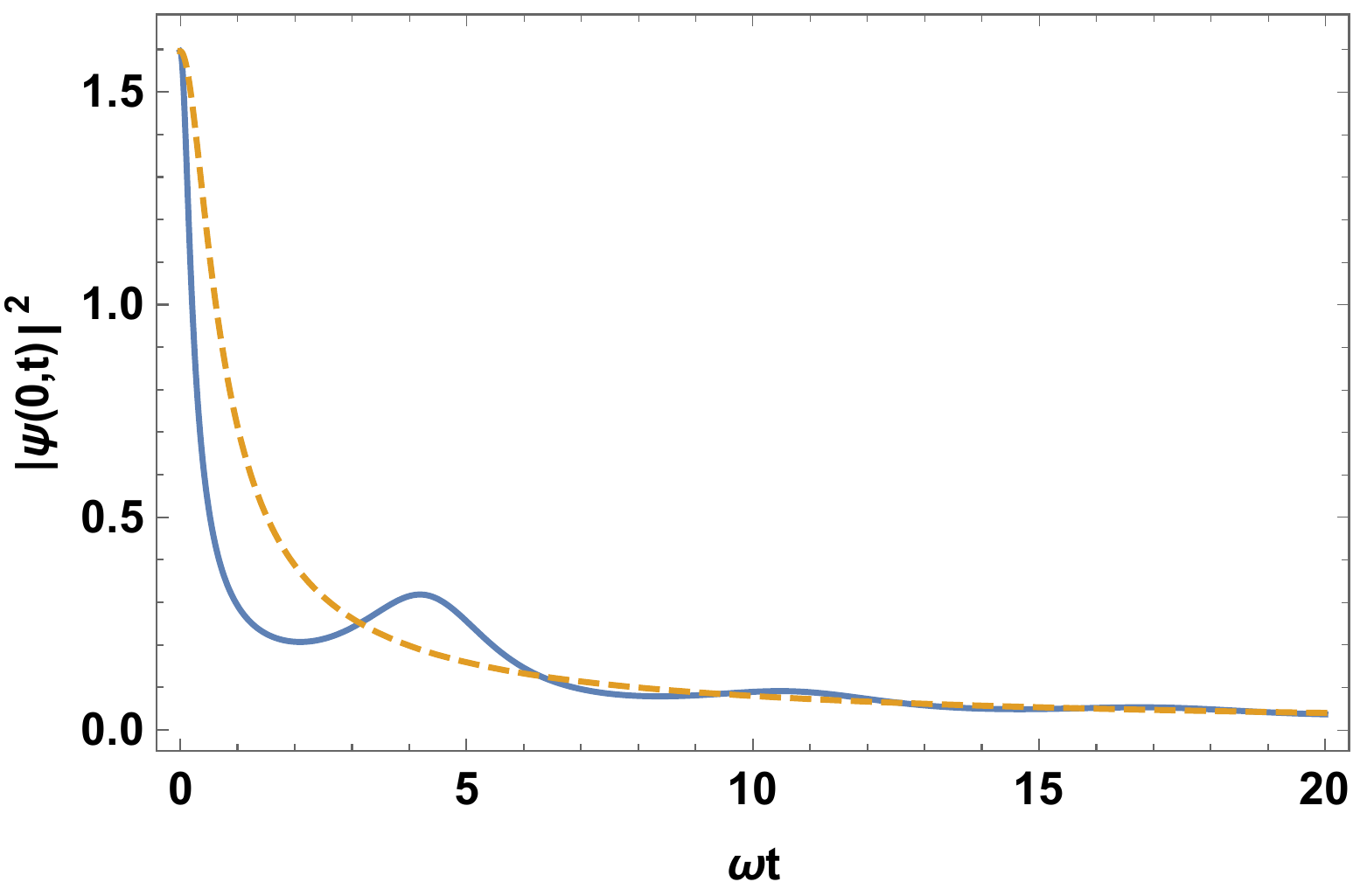}
	\caption{The probability of detecting a particle (at $x=0$) always decreases over time for a free particle in the absence of a GW (dotted line). In comparison, the probability of detecting the same particle can increase in the presence of the GW (solid line). }
	\label{fig:p}
\end{figure}

As a GW can betray the mass of a quantum particle, it is tantalising to ask whether one could use this fact to detect the presence of a GW. As a simple case study, let us consider a particle detector that is at rest in the local co-ordinate system, measuring the probability of a particle being at position $x_0$. At time $t_0$, we place a particle at $x_0$. In the absence of the GW, the probability of detecting the particle always decreases as the wave function spreads according to Schrödinger's equation, as shown by the dotted line in Fig.~\ref{fig:p}. In the presence of the GW, the behaviour of the particle is distinctly different; in particular the probability of detecting the particle at $x_0$ can increase, as shown by the solid line in Fig.~\ref{fig:p}. Therefore, in principle one could simply look for these characteristic increases in the probability of particle detection as signatures of a GW. In practice however, these characteristic increases in the probability of detection is restrictively small for known GW sources. For example, let's consider a rubidium atom ($m=1.4\times 10^{-25}$ kg) in a GW generated by the PSR B1913+16 binary pulse system, which has an amplitude of  $f_0=10^{-14}$ on Earth and frequency $\omega=2.4\times10^{-4}$~\cite{quach16}. To detect this GW one would require a probability detection resolution of $10^{-29}$. With such a small change in the probability distribution, it is unlikely that observing a \textit{single} quantum particle in this manner, can betray the presence of a GW. Some possible avenues for further investigation beyond single particles is provided below.

\section{Outlook}
\label{sec:Outlook}

We have shown that quantum particles violate the WEP in a GW background, in the sense that extra mass information can be extracted in the presence of the GW. We have quantified the degree of violation with the mass Fisher information. We estimated that by observing a single quantum particle, it is unlikely that this violation could be used as a means of GW detection. This however does not preclude more sophisticated setups in utilising quantum WEP violation for  GW detection, such as ensembles of interacting particles~\cite{napolitano11}, high precision measurements with quantum metrology~\cite{vittoria06}, or other novel setups: of note are works that propose that WEP violating Cooper pairs in superconductors may give rise to GW detectors~\cite{papini70,anandan81,chiao82,anandan82,anandan84,anandan85,peng90,peng91}, transducers~\cite{chiao03, licht04}, and mirrors~\cite{minter10,quach15,quach15b}. In these works it is conjectured that the delocalised Cooper pairs behave differently from the localised ion cores of the superconductor in the presence of a GW, due to the larger degree of WEP violation in the former. Our work offers an approach that may be able to prove this conjecture.

\begin{acknowledgements}
This work was financially supported by the Ramsay Fellowship.
\end{acknowledgements}

\bibliographystyle{spphys}
\bibliography{wep}

\begin{thebibliography}{10}
\bibitem{note1}
One notes that cigar shape potential traps could confine a particle to an
  effective one spatial dimension.
\end{thebibliography}

\end{document}